\documentclass[twocolumn,showpacs,preprintnumbers,amsmath,amssymb]{revtex4}
\usepackage{graphicx}% Include figure files
\usepackage{dcolumn}% Align table columns on decimal point
\usepackage{bm}% bold math

\begin{document}

\preprint{APS/123-QED}

\title{Temperature dependent N\'eel wall dynamics in GaMnAs/GaAs\\}% Force line breaks with \\

\author{J. Honolka}%
\email{j.honolka@fkf.mpg.de}
\affiliation{%
Max-Planck-Institut f\"ur
Festk\"orperforschung, Heisenbergstrasse 1, 70569, Stuttgart, Germany}% Lines break automatically or can be forced with \\

\author{L. Herrera Diez}
 \affiliation{%
 Max-Planck-Institut f\"ur
Festk\"orperforschung, Heisenbergstrasse 1, 70569, Stuttgart, Germany}% Lines break automatically or can be forced with \\

\author{R. K. Kremer}%
\affiliation{%
Max-Planck-Institut f\"ur
Festk\"orperforschung, Heisenbergstrasse 1, 70569, Stuttgart, Germany}% Lines break automatically or can be forced with \\

\author{K. Kern}%
\affiliation{%
Max-Planck-Institut f\"ur
Festk\"orperforschung, Heisenbergstrasse 1, 70569, Stuttgart, Germany}% Lines break automatically or can be forced with \\

% \author{A. Enders}%
% \affiliation{%
% Department of Physics and Astronomy, University of Nebraska, Lincoln, NE 68588, USA}% Lines break automatically or can be forced with \\

\author{E. Placidi}%
\affiliation{%
Dipartimento di Fisica, Universit\`a di Roma 'Tor Vergata',
CNR-INFM, Via della Ricerca Scientifica 1, I-00133 Roma, Italy}% Lines break automatically or can be forced with \\

\author{F. Arciprete}%
\affiliation{%
Dipartimento di Fisica, Universit\`a di Roma 'Tor Vergata',
CNR-INFM, Via della Ricerca Scientifica 1, I-00133 Roma, Italy}% Lines break automatically or can be forced with \\

\date{\today}% It is always \today, today,
             %  but any date may be explicitly specified

\begin{abstract}
Extensive Kerr microscopy studies reveal a strongly temperature
dependent domain wall dynamics in Hall-bars made from compressively
strained GaMnAs. Depending on the temperature magnetic charging of
domain walls is observed and nucleation rates depend on the
Hall-geometry with respect to the crystal axes. Above a critical
temperature where a biaxial-to-uniaxial anisotropy transition occurs
a drastic increase of nucleation events is observed. Below this
temperature, the nucleation of domains tends to be rather
insensitive to temperature. This first spatially resolved study of
domain wall dynamics in patterned GaMnAs at variable temperatures
has important implications for potential single domain magneto-logic
devices made from ferromagnetic semiconductors.

\end{abstract}


\maketitle

The ferromagnetic semiconductor GaMnAs\cite{ohno} has been
extensively studied in the past few years not only in the viewpoint
of basic science but also focusing the attention on properties that
can lead to novel applications in spin-based electronics and
magneto-logic devices~\cite{jungwirth,dietl}. For the latter, a good
understanding of domain wall (DW) dynamics is needed in order to
control processes such as the DW nucleation and propagation. In
ferromagnetic GaMnAs with in-plane magnetization, magnetic reversal
processes have been studied mostly by means of
magneto-transport~\cite{tang, pappert}, however with very limited
gain of local information on DW nucleation and motion. Very recently
single DWs have been resolved in the static limit by means of
electron holography on the scale of a few
micrometers~\cite{sugawara08} with high spatial resolution. In
contrast, we have shown that Kerr microscopy provides full time and
spatially resolved information on the dynamics of in-plane magnetic
domains during the magnetization reversal on the scale of a few
hundred micrometers~\cite{ourpaper}.
\newline\noindent Due to the low Curie-temperatures $T_c$ well below room
temperature of most ferromagnetic semiconductors like GaMnAs it is
of technical interest to study these materials in the highest
possible temperature range just below $T_c$. In this work we present
a careful characterization of the temperature dependent biaxial and
uniaxial magnetic anisotropies in compressively strained GaMnAs and
their influence on the evolution of the magnetic domain structure
thereby identifying limits for domain wall logic devices in the high
temperature regime. A preferential DW alignment is found to be
linked to the change in the position of the easy axis given by the
temperature dependence of the uniaxial and biaxial anisotropy
contributions. An increase in the number of domain nucleation
centers is observed beyond a critical temperature where a
biaxial-to-uniaxial anisotropy transition takes place. The
dependence of this behaviour on the geometry of the device
is also presented.\\
The material under study, consists of GaMnAs epilayers of 170 nm
thickness grown on GaAs(001) by molecular beam epitaxy (MBE). The
nominal Mn concentration is (2.3 $\pm$ 0.1)\%  and has been
estimated on the bases of flux ratios. A more detailed description
of the sample growth and material characterization has been given
elsewhere~\cite{ourpaper}. The GaMnAs devices used in the Kerr
microscopy experiments are Hall bars of 200$\mu$m width fabricated
by standard photolithography and ion milling.

\section*{Magnetic characterization of unpatterened, virgin GaMnAs epilayers}

\begin{figure}
  % Requires \usepackage{graphicx}
  \includegraphics[width=9cm, clip,angle=0]{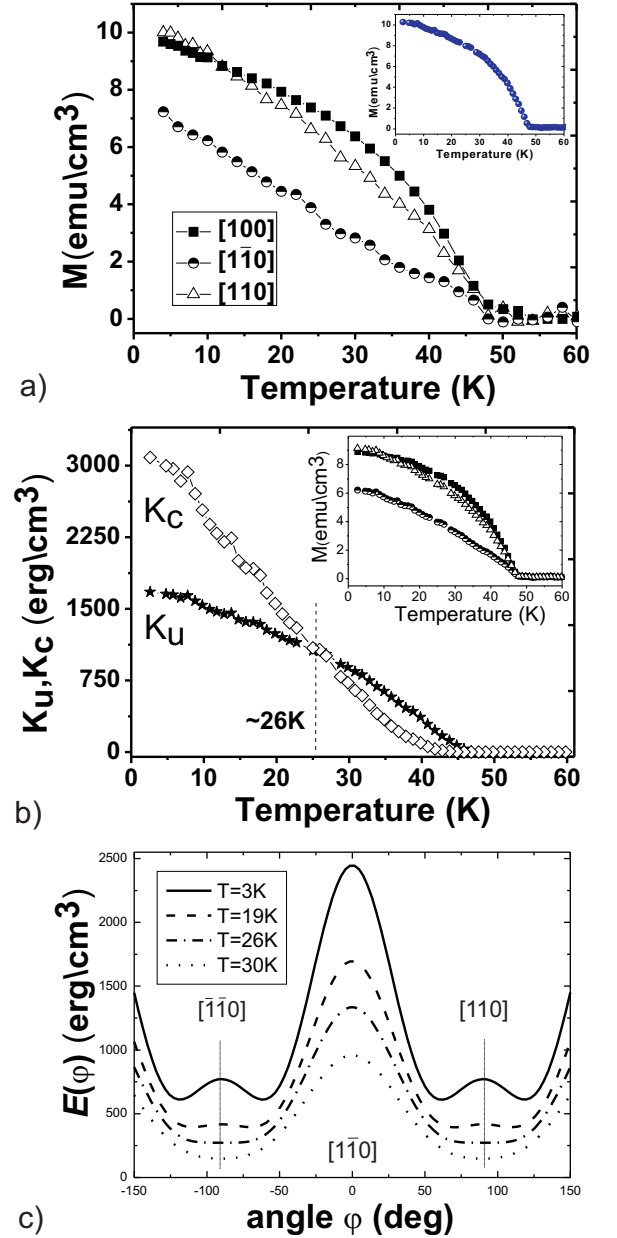}\\
    \vspace{-25pt}
  \caption{(a) Temperature dependent SQUID measurements along [1$\bar{1}$0], [110] and
  [100] in a field of 50Oe. The inset shows the
  magnetization at a saturating field of 1 Tesla. (b) Temperature dependence of $K_{u}$ and
  $K_{c}$. The values have been derived by fitting the data shown in (a) using the saturated
  SQUID magnetization data and assuming Stoner-Wohlfahrt behavior and magnetization dependent
 anisotropy constants $K_{u}$= 30.0$M^{1.8}$ and $K_{c}$=0.32$M^{4.1}$. The fits for all
 directions are shown in the inset. (c) Energy density in the absence of a magnetic field for
 different temperatures. The plots are generated using the measured values for
 $K_{u}(T)$, $K_{c}(T)$ and of the magnetization $M(T)$.}\label{squid}
\end{figure}

For a full characterization of the magnetic anisotropy within the
GaMnAs epilayer we performed temperature dependent SQUID as well as
magneto-optical Kerr effect (MOKE) measurements with magnetic fields
applied in various in-plane directions. SQUID measurements were
performed cooling the sample in a field of 1000 Oe and applying a
field of $50$Oe during the measurement. The results are shown with
in Fig.~\ref{squid}(a) for fields along three directions
[1$\bar{1}$0], [110] and [100]. Also plotted is the magnetization
versus temperature $M(T)$ (Fig.~\ref{squid}(a), inset) in a
saturating field of $H=1$T. From the temperature dependent magnetic
response at non-saturating fields of $H=50$Oe for different
directions the temperature dependence of the anisotropy constants
can be estimated assuming a Stoner-Wohlfahrt coherent rotation of
the magnetization following the total energy density
$E(\varphi)=\frac{K_{c}}{4}\cos^{2}(2\varphi) + K_{u}\cos^{2}\varphi
- M H\cos(\varphi-\varphi_{H})$, where $K_{c}$ and $K_{u}$ are the
biaxial and uniaxial anisotropy constants, $\emph{M}$ is the
magnetization, $\emph{H}$ the magnetic field, and $ \varphi$ and
$\varphi_{H}$ are the angles of $\emph{M}$ and $\emph{H}$ with the
[1$\bar{1}$0] direction. For each temperature the measured SQUID
signal $M^{\text{SQUID}}$ is determined simply by the equation
system
\begin{eqnarray}
\label{energy} {\partial}E/{\partial}\varphi=0, ({\partial^2}E/{\partial^2}\varphi>0)\\
M^{\text{SQUID}} = M \cos(\varphi-\varphi_{H})
\end{eqnarray}
% --------------------------------------------------------
% --------------------------------------------------------
Here $M^{\text{SQUID}}$ is the measured projection of the
magnetization $M(T)$ on the axis of the SQUID pick-up coils, which
are aligned parallel to the magnetic field. While $M(T)$ is known
from the SQUID measurement at saturating fields, $K_{u}$ and $K_{c}$
are temperature dependent parameters to be derived by fitting.
Assuming a magnetization dependence of the anisotropy constants
close to $K_{u}=\alpha M^{2}$ and $K_{c}= \beta M^{4}$~\cite{wang}
we can use equations (1) and (2) to fit the SQUID data as shown in
the inset of Fig.~\ref{squid}(b). The fits shown for the three
directions [1$\bar{1}$0], [110] and [100] are derived using one and
the same fit parameters $\alpha= 30.0$ and $\beta= 0.32$ in addition
to the magnetization exponents 1.8 and 4.1 for the expressions of
$K_{u}$ and $K_{c}$, respectively.
% ------------------------------------------------
\begin{figure*}
  % Requires \usepackage{graphicx}
  \includegraphics[width=18cm]{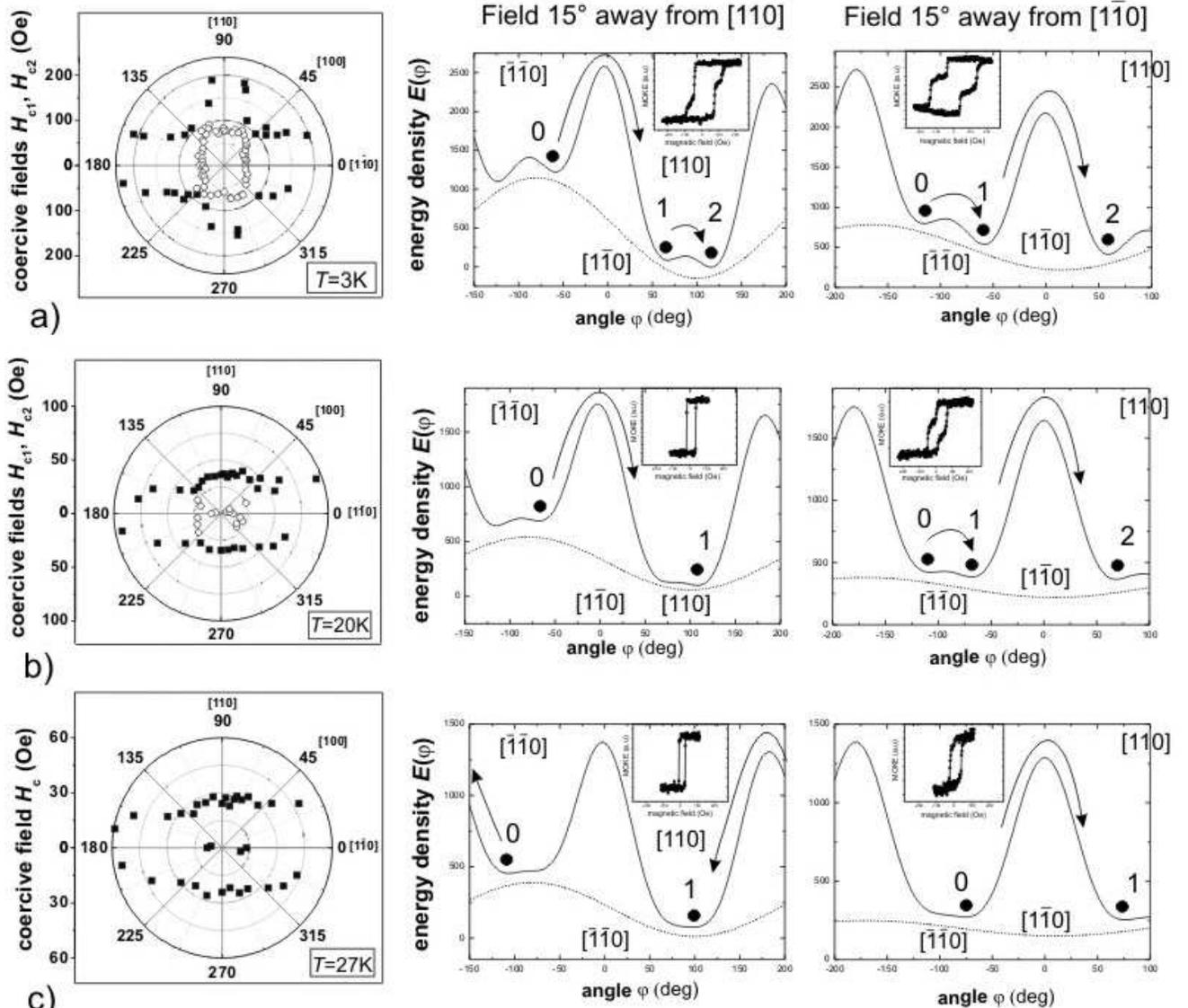}\\
  \vspace{-15pt}
  \caption{The polar plots (a), (b) and (c) to the left show the measured coercivities
  as a function of the angle $\varphi_{H}$ of the applied field with respect to the [1$\bar{1}$0] direction
  at temperatures of 3K, 20K, and 27K, respectively. At low temperatures two coercive fields
  $H_{c1}$ and $H_{c2}$ appear while at temperatures $T>26$K the entire magnetic transition happens at
one single field $H_{c}$. The two diagrams right of each of the
three polar plots show the energy density $E(\varphi)$ at fields
$H_{c1}(T)$ applied $15^{\circ}$ away from the
[110]($\varphi_{H}=105^{\circ}$) and
[1$\bar{1}$0]($\varphi_{H}=15^{\circ}$) directions, respectively.
The insets show the MOKE hysteresis measurement at the respective
angles. As a reference the Zeeman energy term is also plotted in
dashed lines. The energy densities are plotted using the temperature
dependent values for $K_{u}$, $K_{c}$ and the magnetization
$M$.}\label{polar}
\end{figure*}
% --------------------------------------------------
In Fig.~\ref{squid}(b) the temperature dependence of $K_{u}$ and
$K_{c}$ is plotted as a result of the fitting procedure. A clear
crossover is observed from biaxial to uniaxial magnetic anisotropy
at approximately 26K where $K_{u}=K_{c}$. As a consequence along the
[110] direction the second derivative of the energy,
${\partial^2}E/{\partial^2}\varphi$, changes sign at $K_{u}=K_{c}$
and the number of local minima in $E(\varphi)$ is reduced from 4 to
2 due to the disappearance of the biaxial induced energy barrier in
the [110] direction (see Fig.~\ref{squid}(c)). As extensively shown
in magneto-transport measurements by Pappert {\it et
al.}~\cite{pappert} this crossover is directly visible in polar
coercivity plots of Fig.~\ref{polar}, which summarize the coercive
fields derived from MOKE hysteresis loops taken in different
directions with respect to the [1$\bar{1}$0] crystal axis. The shape
of the angular dependence of the coercivities at $T=3$K and $T=27$K
clearly confirms the change from a four-fold $K_{c}$ dominated to a
two-fold $K_{u}$ dominated symmetry at low and high temperatures,
respectively. At low temperatures in agreement with the literature
the biaxial four-fold symmetry leads to two step reversals via
intermediate local minima in $E(\varphi)$. Specifically for our
samples transitions at $T=3$K have been shown to be mediated by two
individual domain walls with DW angles $\Delta\varphi
\sim$120$^{\circ}$ and $\sim$60$^{\circ}$,
respectively~\cite{ourpaper}, triggered at the coercive fields
$H_{c1}$ and $H_{c2}$. The reversal via an intermediate state is
illustrated in the two right plots of Fig.~\ref{polar}(a) where
$E(\varphi)$ is shown for fields $H=H_{c1}$ applied along angles
$15^{\circ}$ away from the [110] and [1$\bar{1}$0] directions,
respectively. For $H_{c1}$ the measured values at $T=3$K in the
respective direction were taken. From the diagram it also becomes
immediately clear that we expect $H_{c1}$ measured close to the
[110] directions to be higher compared to [1$\bar{1}$0] because in
the former case the barrier that has to be overcome is governed by
the larger uniaxial part $K_{u}$ of the anisotropy landscape. Since
we will later characterize DW transitions in detail using Kerr
microscopy we want to stress the fact that measuring the
coercivities at different $\varphi_{H}$ can trigger magnetization
transitions with either clockwise (CW) or counter-clockwise (CCW)
sense of rotation. From Fe/GaAs thin film systems with an equivalent
magnetic anisotropy symmetry it is known that the sense of rotation
changes whenever the magnetic field direction $\varphi_{H}$ crosses
a local minimum or a maximum in the magnetic energy landscape
$E(\varphi)$~\cite{Daboo95}.
% --------------------------------------------------------
\begin{figure}
  % Requires \usepackage{graphicx}
  \includegraphics[width=9cm, clip,angle=0]{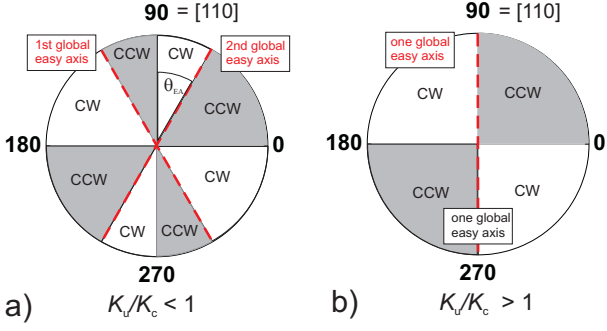}\\
    \vspace{-5pt}
  \caption{Magnetic field direction ($\varphi_{H}$) dependent change of the sense of the magnetization
  reversal process for (a) $K_{u}/K_{c}<1$ and (b) $K_{u}/K_{c}>1$. White areas represent
  clockwise (CW) rotation, grey areas counterclockwise (CCW).}\label{DWsymmetry}
\end{figure}
% --------------------------------------------------------
Therefore, as shown in Fig.~\ref{DWsymmetry}(a), at low temperatures
where $K_{u}/K_{c}<1$, the sense of the transition changes 8 times
when $\varphi_{H}$ is swept over the full angle range~\footnote{The
change of the sense at the maxima of $E(\varphi)$ at [1$\bar{1}$0]
and [110] is directly visible when measuring the magneto-transport
in our Hall-bars. The transverse Hall-voltage $V_{\text xy}$ changes
sign for measurements $\pm 5^{\circ}$ away from the respective
directions.}. Four of the eight sign changes occur when the magnetic
field direction crosses the two equivalent global easy axis
directions located at angles $\pm \theta_{\text{EA}}(T)$ away from
the [110] direction as shown in Fig.~\ref{DWsymmetry}(a). At high
temperatures $T > 26$K in agreement with Fig.~\ref{polar}(c) only
one single transition at $H_{c}$ is observed and we expect DWs with
angles $\sim$180$^{\circ}$. Accordingly, we expect that the sense of
the transitions only change sign 4 times during a full angle sweep
of $\varphi_{H}$ (see Fig.~\ref{DWsymmetry}(b) for the case
$K_{u}/K_{c}>1$). At $T=27$K in a very narrow angle window close to
the [1$\bar{1}$0] low coercivities of about $15$Oe are found. When
the field direction sufficiently deviates from the [1$\bar{1}$0]
axis, coercivities quickly jump to higher values larger than
$30$~Oe. In an angle window $\pm 30^{\circ}$ away from the [110]
direction the values of the coercivities are stable around
$\sim$30~Oe indicating that in this region the magnetic reversal is
highly reproducible and not critically dependent on the sample
orientation. Therefore, applying the field along the [110] direction
at different temperatures below and beyond the crossing point of
$K_{u}$ and $K_{c}$ should allow for the observation of the
transition between $\sim$120$^{\circ}$ and $\sim$180$^{\circ}$ DWs.
From an application point of view this direction is interesting
since in this regime the transition was shown to be propagation
dominated with a relatively small number of domains involved in the
process~\cite{ourpaper}.

\section*{Observation of temperature dependent domain wall dynamics in patterned GaMnAs Hall-bars}

\begin{figure}
  % Requires \usepackage{graphicx}
  \includegraphics[width=8cm]{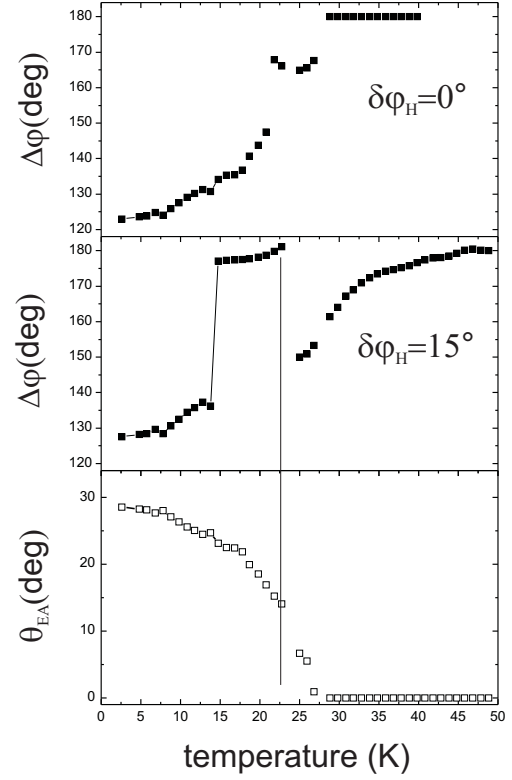}\\
  \vspace{-5pt}
  \caption{Plot of the angle between the [110] axis and the closest
  global minimum direction versus temperature (open squares). The
  global easy axis reaches the [110] direction at about 26K where $K_{u} =
  K_{c}$ (compare to Fig.\ref{squid}).
  The corresponding DW angle for a transition via the uniaxial easy axis along
  [110] and with a deviation of $\delta\varphi_{H}=15^{\circ}$ is also plotted (filled symbols).
  The DW angle increases monotonously from $\sim120^{\circ}$ and reaches 180$^{\circ}$ at higher
  temperatures.}\label{angle}
\end{figure}

The Kerr-microscopic observation of magnetic domains was performed
using the same procedure as described in Ref.~\cite{ourpaper}.
Before presenting the microscopy results which focus on the
temperature dependent dynamics of DWs for magnetic fields applied
close to the [110] direction we would like to shortly discuss the
expected change of the DW angle $\Delta\varphi$ and sense of
rotation with temperature as well as with increasing deviations
$\delta\varphi_{H}$ from the [110] direction.\newline From simple
symmetry arguments reflected in Fig.~\ref{DWsymmetry} it is evident
that generally small deviations of $\pm\delta\varphi_{H}$ to both
sides of the [110] direction will trigger DW transitions of opposite
sense. However, despite the opposite sense in rotation the absolute
DW angles remain exactly the same. More specifically, at low
temperatures $T<26$K and $\delta\varphi_{H}<\theta_{\text{EA}}$ CW
(CCW) deviations lead to CW (CCW) transitions at $H_{c1}$ and
$H_{c2}$, whereas for $\delta\varphi_{H}>\theta_{\text{EA}}$ CW
(CCW) deviations lead to a CCW (CW) transition. The angle
$\theta_{\text{EA}}$ is shown in Fig.~\ref{DWsymmetry}(a). For
$T>26$K CW (CCW) deviations always lead to a CCW (CW) transition.
For a full understanding of DW dynamics at different temperatures it
is therefore important to trace the temperature dependent global
easy axis direction. To give an example of the influence of
$\delta\varphi_{H}$ on $\Delta\varphi$, Fig.~\ref{angle} shows the
temperature dependent angle $\theta_{\text{EA}}(T)$ of the global
easy axis direction with respect to [110] at zero magnetic field
together with the expected DW angle of the first transition at
$H_{c1}$ for $\delta\varphi_{H}=0^{\circ}$ and for a field deviation
$\delta\varphi_{H}=15^{\circ}$ with respect to [110]. The easy axis
directions were obtained by tracing one of the two energy minima in
$E(\varphi)$ closest to the [110] uniaxial easy axis (see
Fig.~\ref{squid}(c)). $\theta_{\text{EA}}(T)$ is determined by
$K_{u}(T)$ and $K_{c}(T)$ given in Fig.~\ref{squid}. As expected at
$\sim$26 Kelvin, the temperature of the crossing between $K_{u}$ and
$K_{c}$ (see Fig.~\ref{squid}), the global easy axis starts to be
fully aligned with the [110] direction. The calculation of the DW
angle of the first transition includes the temperature dependence of
the coercive field applied close to the [110] direction. Coherent
rotation effects in two domains separated by the DW are thus taken
into account.
% -------------------------------
% -----------------------------------------
The DW angle $\Delta\varphi(T)$ for $\delta\varphi_{H}=0^{\circ}$
and for the field deviation $\delta\varphi_{H}=15^{\circ}$ shows two
distinct jumps caused by the sequential destabilization of the
initial and the final magnetization state of the transition. In the
three middle row plots of Fig.~\ref{polar} this effect is
illustrated for the case $\delta\varphi_{H}=15^{\circ}$. At around
20K the intermediate state of the CCW two-step transition becomes
instable and the magnetization rotates to the final state in one
single step (in the MOKE hysteresis shown in the inset the
intermediate step has vanished). $\Delta\varphi$ thus increases
abruptly at this point, however the CCW sense of the transitions is
preserved. Finally, at temperatures $T\sim23$K where
$\theta_{\text{EA}}=\delta\varphi_{H}=15^{\circ}$ (see
Fig.~\ref{angle}) the sense of the transition changes to CW and the
initial magnetization state rotates towards the [$\bar{1}\bar{1}$0]
direction leading to a reduction in $\Delta\varphi$. As the
temperature is further elevated
the initial and final magnetization states approach the global easy axis direction along [110].\\
In the following the effect of the temperature dependent change from
four-fold to two-fold symmetry in $E(\varphi)$ on the DW dynamics is
studied on the basis of extensive Kerr microscopy measurements with
magnetic fields applied close to the [110] direction.

\subsection*{Domain wall alignment - Charging of walls}

\begin{figure}
 %  Requires \usepackage{graphicx}
 \includegraphics[width=8cm]{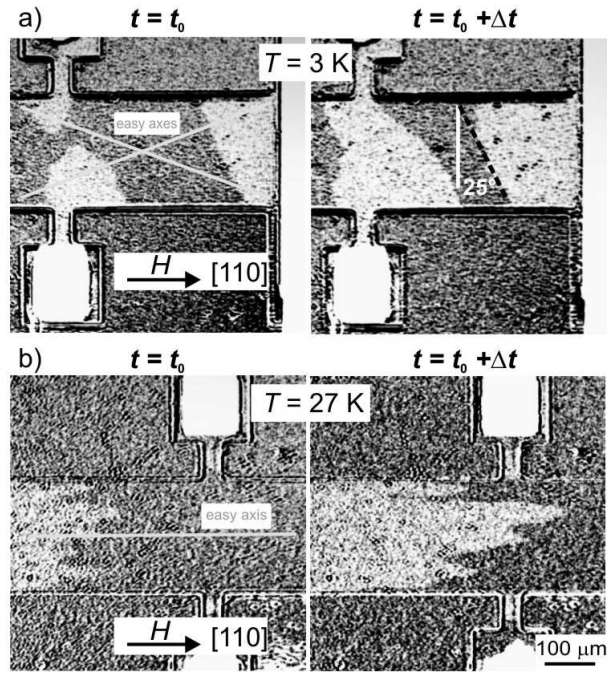}\\
 \caption{Kerr images of the domain structure at 3 Kelvin (a) and 27 Kelvin (b) (upper and lower plots, respectively)
 in a Hall-bar oriented along [110]. Left and right images are consecutive frames taken at times $t=t_0$ and
$t=t_0+\Delta t$ to picture the time evolution. The magnetic field is applied along [110]. The DW orientation changes
significantly with temperature while the number of nucleation centers is not strongly affected.
Easy axes directions are indicated by green lines.}\label{alignment}
\end{figure}

\begin{figure}
 %  Requires \usepackage{graphicx}
 \includegraphics[width=8cm]{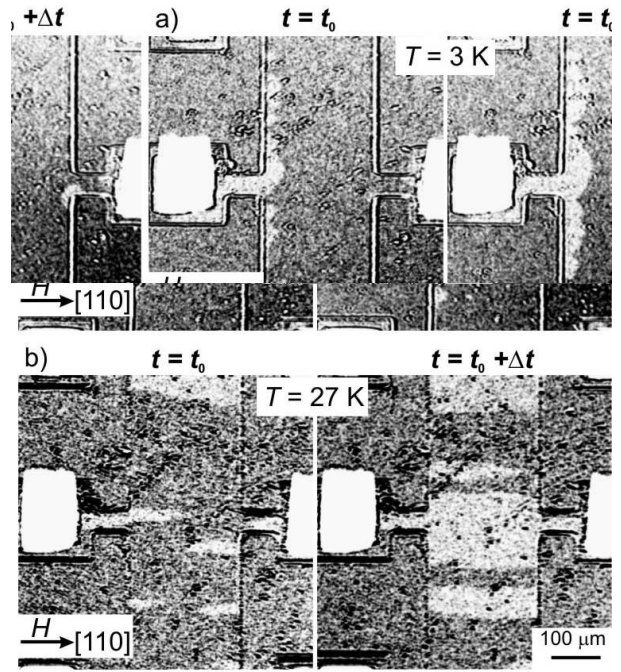}\\
 \caption{Kerr images of the domain structure under the same measuring conditions as for those
shown in Fig.~\ref{alignment}, however with the Hall-bar oriented
along [1$\bar{1}$0].}\label{alignment2}
\end{figure}

The Kerr images in Fig.~\ref{alignment}(a) and (b) (Hall-bar
$\|$~[110]) and Fig.~\ref{alignment2}(a) and (b) (Hall-bar
$\|$~[1$\bar{1}$0]) show typical domain structures for the field
applied along the [110] direction at 3K and 27K, respectively. For
all four cases two consecutive frames at times $t=t_0$ and
$t=t_0+\Delta t$ were extracted from a movie to picture the time
evolution. Fig.~\ref{alignment} and Fig.~\ref{alignment2}
demonstrate that the alignment of the DWs with respect to the [110]
direction is clearly temperature dependent. While at low
temperatures the DWs avoid the alignment with the [110] direction
along the Hall bar they prefer the parallel alignment at higher
temperatures in both cases. Only the DW nucleation behavior seems to
be dependent on the Hall-bar orientation. Here we observe that only
in Fig.~\ref{alignment2} nucleation happens preferentially at the
long sides of the Hall-bar. We will discuss nucleation effects in
detail in the next section.
% -----------------------------------------
% -----------------------------------------
% -----------------------------------------
% -----------------------------------------
\newline\noindent In most magnetic systems the alignment of DWs is correlated
to the surface divergence of the magnetization at the domain
boundary due to magnetization components normal to the
DW~\cite{bookdomains,bookred,bookblue}. In general, this creates
so-called magnetic charges proportional to
$(\bf{M}_{1}-\bf{M}_{2})\cdot \hat{\bf{n}}$ at the DW boundary
accompanied by a cost of stray field energy, where $\bf{M}_{1}$ and
$\bf{M}_{2}$ are the magnetization vectors of the two domains
separated by the DW and $\hat{\bf{n}}$ the wall normal facing
towards domain 2. Hence, in our GaMnAs samples, in order to avoid
magnetic charges, DWs should be aligned along [110] for the low
temperature $\sim$120$^{\circ}$ DW transition with fields along
[110]. At higher temperatures in the case of $180^{\circ}$ DW
transitions where the magnetization vectors $\bf{M}_{1}$ and
$\bf{M}_{2}$ are collinear with the global easy axis along [110] we
expect the system to try and avoid head-to-head type of boundaries
$\hat{\bf{n}}\| [110]$ with maximum amounts of magnetic
charges~\cite{vogel}. In agreement with the latter, the observed DWs
at 27 K show typical zigzag patterns throughout the reversal
dynamics with $\hat{\bf{n}}$ pointing preferentially parallel to the
[1$\bar{1}$0] direction where $(\bf{M}_{1}-\bf{M}_{2})\cdot \hat{n}
= 0$ holds. At low temperatures, however, out results are clearly
not according to the above described model. As discussed, the
reversal dynamics shown in Fig.~\ref{alignment} (a) with the field
applied along the [110] direction corresponds to a
$\sim$120$^{\circ}$ DW~\cite{ourpaper} where the initial and final
magnetization states $\bf{M}_{1}$ and $\bf{M}_{2}$ are aligned with
two of the biaxial global easy axes which lay at $\sim\pm25^{\circ}$
from the [110] direction (see Fig.~\ref{angle}). The preferential
orientation of the DWs around 25$^{\circ}$ away from [1$\bar{1}$0]
observed in the Kerr images, thus, points to significant amounts of
magnetic charges $\sim M \text{cos}(25^{\circ})$ accumulated at the
walls.
\begin{figure}
  % Requires \usepackage{graphicx}
  \includegraphics[width=8cm,clip,angle=0]{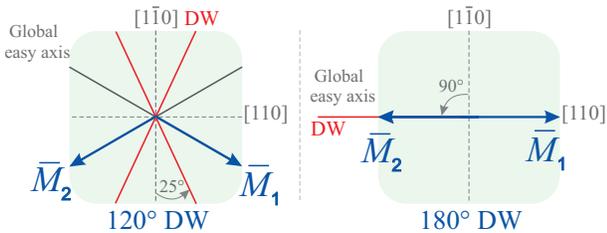}\\
  \vspace{-5pt}
  \caption{Diagram of the magnetization reversal process and observed
   DW orientation (red lines) at 3 Kelvin (a) and 27 Kelvin (b).
    The global easy axes are indicated by full gray lines, and [110] and [1$\bar{1}$0]
    crystal directions by dashed lines. The orientation of 180$^{\circ}$ DWs
    at higher temperatures is along the easy axis [110](90$^{\circ}$ away from [1$\bar{1}$0]) while 120$^{\circ}$
    DWs orient preferentially around 25$^{\circ}$ away from [1$\bar{1}$0].}\label{diagram}
\end{figure}
Fig.~\ref{diagram}(left) and (right) summarizes the experimentally
observed DW orientations for the case of 120$^{\circ}$ and
180$^{\circ}$ DW transitions at low and high temperatures together
with the respective easy axis directions. It should be noted at this
point that the DW orientations are found to be the same in our
virgin film samples and therefore are not a consequence of the
Hall-bar patterning process.\newline

Before we start to discuss the physics leading to the
observed DW alignment behavior it is helpful to estimate the
expected contributions of the stray field to the energy density. In
the diluted ferromagnetic semiconductor GaMnAs the magnetization is
about 2 orders of magnitude lower compared to typical 3$d$ metal
ferromagnets like Fe and therefore stray field energy contributions
to the total energy density are generally reduced by a factor
10$^{-4}$. One of the consequences is that in GaMnAs films N\'{e}el
walls are energetically preferred to Bloch walls up to relatively
large film thicknesses $d_{\text{crit}}$. In this regime the system
avoids magnetic surface charges at the film surfaces and encounters
volume charges within the N\'{e}el wall. It was shown that the
critical thickness can be approximated by $d_{\text{crit}}=
13.8\sqrt{A/4\pi M^2}$, where $A$ is the exchange coupling constant.
With a typical value of $A=4\times 10^{-8}\text{erg}/\text{cm}$ for
GaMnAs and magnetization values of the order of
10~$\text{emu}/\text{cm}^3$ one gets a critical thickness of about
$1\mu$m (compare to Permalloy where $d_{\text{crit}}$=50nm). Thus,
we can assume that the magnetic dynamics in our GaMnAs films of
170nm thickness is governed by N\'{e}el-type walls in agreement with
experiments by Sugawara {\it{et al.}}.~\cite{sugawara08}.\\
The low magnetization value also reduces the stray field energy
density $\epsilon_S$ caused by magnetic charges situated at a DW. It
is given by $\epsilon_S =2\pi(\bf{M}_{1}-\bf{M}_{2})\cdot
\hat{n})^2$ for an infinitely extended DW. For epitaxial Fe films of
$150{\text \AA}$ thickness grown on GaAs with a predominant cubic
anisotropy $K_{c}>0$ and large values $\epsilon_S$ of the order of
1$\times 10^{6}\text{erg}/\text{cm}^3$ a strict preferential DW
alignment according to the stray field minimization condition
$\epsilon_S = 0$ has been reported by Gu {\it et al.}~\cite{gu}. The
authors observe the alignment of 90$^{\circ}$ and 180$^{\circ}$ DWs
with the hard and easy axis, respectively, when the field is applied
along the easy axis parallel to the cubic crystal symmetry
direction. For our GaMnAs samples with low concentrations of Mn,
however, stray field energy densities $\epsilon_S$ are only of the
order of 100~$\text{erg}/\text{cm}^3$ at most. Moreover, in thin
films $\epsilon_S$ is further reduced due to the limited lateral
extension of the DW when oriented perpendicular to the film. It can
be shown that stray fields produced by magnetic charges in laterally
confined N\'{e}el walls decay like $1/x^2$ at large distances away
from the wall~\cite{Kronmueller65}. As a consequence for very thin
films more complex N\'{e}el wall shapes occur and total wall
energies have to be evaluated numerically including exchange
stiffness and magnetic anisotropy, which leads to solutions
including isolated charged walls with $\epsilon_S \neq
0$~\cite{hubert79}. From calculations by A. Hubert~\cite{hubert79}
with therein defined dimensionless parameters $Q=K/2\pi M^2$ and
$\lambda = 2Q\sqrt{A/K}/d$ one expects charged 120$^{\circ}$ DWs in
our GaMnAs samples of thickness $d=170$nm with $Q\approx 1$ and
$\lambda \approx 1$ in accordance with our results of the
Kerr-measurements at $T=3K$. DW charging effects similar to ours are
visible also in epitaxial Fe films in the ultrathin film limit grown
on GaAs. Although the authors of Ref.~\cite{gu} did not discuss this
aspect in detail the film thickness dependent cross-over from
uncharged to partly charged 90$^{\circ}$ DWs clearly shows in their
Lorentz microscopy data for $d=150{\text \AA}$ and $d=35{\text
\AA}$, respectively~\cite{Daboo95}. In GaMnAs epilayers Sugawara
{\it{et al.}} found both 90$^{\circ}$ N\'{e}el walls oriented along
the [1$\bar{1}$0] direction and 20$^{\circ}$ away from the
[1$\bar{1}$0] (see DW (iii) in Fig. 1(b) of Ref.~\cite{sugawara08}).
The latter configuration again should correspond to a charged wall
although the authors did not comment on this issue. It should be
noted, however, that since the Lorentz microscopy technique only
permits the observation of domains close to the film edges where a
non-magnetic reference signal is available, the local orientation of
the DWs can also be affected by inhomogeneous morphology induced by
the lithography process as well as flux closure processes.

\subsection*{Temperature dependence of domain nucleation}

As mentioned in the previous section a clear asymmetry in the
nucleation behavior is observed for Hall-bars oriented in the [110]
and [1$\bar{1}$0] direction (see Fig.~\ref{alignment} and
Fig.~\ref{alignment2}). While in the former case, for both
120$^{\circ}$($T=3$K) and 180$^{\circ}$($T=27$K) DWs, a small number
of domains occur and the contact pads of the Hall bar devices tend
to serve as nucleation centers, for [1$\bar{1}$0] oriented Hall-bars
nucleation events happen preferentially at the long sides of the bar
and appear larger in number.\newline
% --------------------------------------------------------
\begin{figure}
  % Requires \usepackage{graphicx}
  \includegraphics[width=8cm]{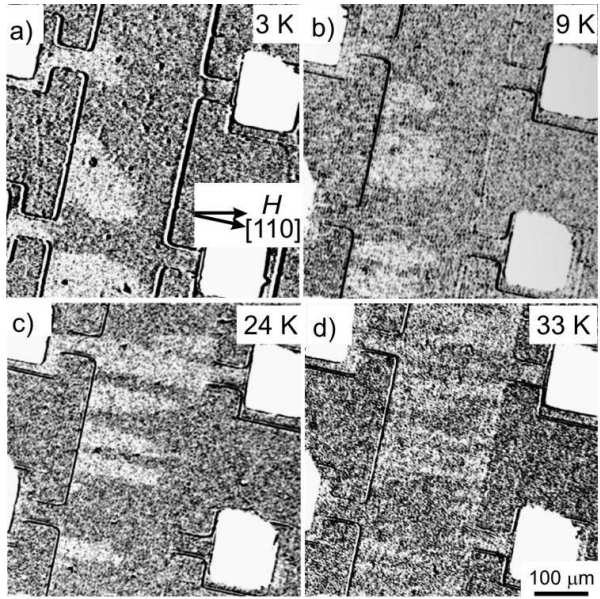}\\
  \caption{Kerr images of the domain structure for the field
  applied $\sim$ 15$^{\circ}$ away from [110] at 3K (a), 16K (b), 24K (c), and 33K (d). }\label{domains}
\end{figure}
% ---------------------------------------------------------
In order to study influences of geometry and temperature on the
nucleation in more detail, Kerr microscopy was performed in small
temperature steps on a Hall-bar having its longitudinal axis along
the [1$\bar{1}$0] axis. We observe that below temperatures of $\sim
24$ Kelvin the number of domains involved in the transition remain
fairly small and constant as shown in Fig.~\ref{domains} (a) and (b)
corresponding to temperatures of 3 and 16 Kelvin, respectively.
However, beyond this temperature the number of nucleation events at
the long sides of the Hall-bar edge grow dramatically and at the
same time domains become increasingly elongated as illustrated in
the Kerr images in Fig.~\ref{domains}(c) and (d) taken at 24 and 33
Kelvin, respectively. The number of domains involved in the reversal
process versus temperature are plotted Fig.~\ref{number} (open
symbols) in a temperature range going from 3 to 33 Kelvin and show
an exponential behavior. Due to the decreasing contrast in the Kerr
signal with decreasing magnetization values Kerr images could not be
evaluated
in the temperature range between 33 Kelvin and $T_c$.\\
% --------------------------------------------------------
\begin{figure}
  % Requires \usepackage{graphicx}
  \includegraphics[width=8cm]{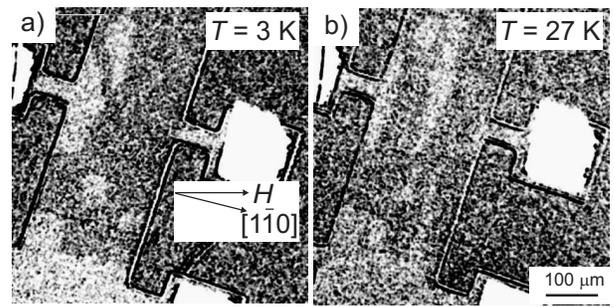}\\
  \caption{Kerr images of the domain structure for the field
  applied $\sim$ 20$^{\circ}$ away from [1$\bar{1}$0] at 3K (a) and 27K (b).}\label{domains2}
\end{figure}
% ---------------------------------------------------------
% --------------------------------------------------------
\begin{figure}
  \vspace{-5pt}
  % Requires \usepackage{graphicx}
  \includegraphics[width=9cm, clip,angle=0]{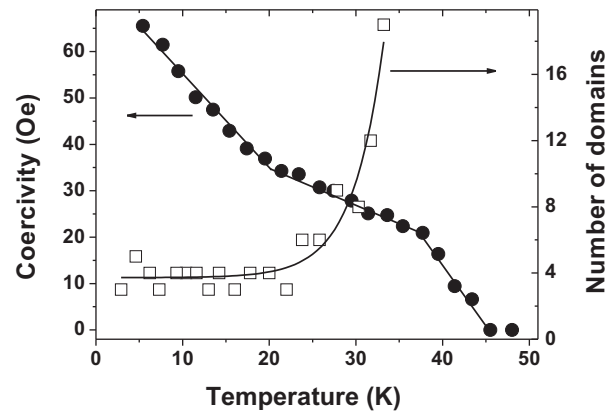}\\
    \vspace{-25pt}
  \caption{Temperature dependence of the coercive fields $H_c$ along the [110] direction and number of domains
  involved in the magnetic reversal process, respectively.}\label{number}
\end{figure}
% --------------------------------------------------------
Comparing the nucleation dynamics for the Hall-bar oriented in the
[110] and [1$\bar{1}$0] directions it is evident that only
$\sim$180$^{\circ}$ DWs appearing at temperatures around 25 Kelvin
are strongly affected by the orientation of the Hall bar with
respect to the crystal axis and field vector. Generally, the Kerr
images in Fig.~\ref{domains} confirm that for fields applied close
to the [110] direction nucleation of domains is happening at film
edges facing the [110] direction. Indeed, Fig.~\ref{alignment} (a)
and (b) show that domains are never nucleated at the edges facing
exactly the [1$\bar{1}$0] direction. Instead nucleation in
Fig.~\ref{alignment} happens at the square shaped Hall-bar pads with
two edges facing [110] or at the far Hall-bar ends (not visible in
the Kerr images) again facing [110]. To prove that the asymmetry is
indeed connected with the crystal orientation we looked at a
Hall-bar patterned in the [110] direction with an applied field
close to the [1$\bar{1}$0] direction (see Fig.~\ref{domains2}). In
accordance to our earlier work~\cite{ourpaper} we see multiple
nucleation events within the film at low temperatures characteristic
for $\sim60^{\circ}$ DW transitions and no preferential nucleation
at the sides of the Hall-bar. At high temperatures $T>27$K again the
domains are elongated along the easy axis direction [110], however
this time preferential nucleation at the long sides of the Hall-bar
is not observed.\newline

Anisotropic nucleation of domains in thin ferromagnetic films as
observed in Fig.~\ref{domains}, where the observed preferred
nucleation occurs at the Hall-bar sides $\|$~[1$\bar{1}$0], can have
different origins:\\

{\it Lithography induced Anisotropies}: During the Hall-bar
lithography process differences in the edge profiles along [110] and
[1$\bar{1}$0] can be introduced. As an example it is known that the
wet etching process of GaAs exhibits a different dynamics in the
respective directions, leading to different edge profiles. However,
the ion milling technique used in our case leads to direction
independent processing and a homogeneous edge profile in all
directions of the Hall-bar. This was verified using x-sectional
scanning electron microscopy. Also lattice relaxation effects as
observed at stripe edges~\cite{Wunderlich, Wenisch} that lead to
local changes in the magnetic anisotropy energy $E(\varphi)$ should
be equal in strength for edges $\|$~[110] and $\|$~[1$\bar{1}$0]. We
therefore claim the observed asymmetric nucleation behavior not to
be a consequence of the Hall-bar patterning process.\\

{\it Anisotropies through closure domains}: Anisotropies in the
nucleation rates can be induced by local dipolar fields, which decay
like $1/x$ away from the edges and trigger flux closure domains. In
micropatterned biaxial epitaxial Fe films on GaAs DW transitions of
90$^{\circ}$ type are triggered preferentially at film edges where
the rotation of the magnetization due to local dipolar fields has
the same sense as the DW transition itself~\cite{Ebels97}. The local
rotation of $M$ can then be understood as a partial transition due
to dipolar fields which facilitates the domain nucleation induced by
the external field $H$. Indeed for the CCW 120$^{\circ}$ DW
transitions shown in Fig.~\ref{domains} ($H$-field that induces the
transition is slightly rotated CW from the global easy axis
direction) we see nucleation at the edges $\|$~[1$\bar{1}$0] where
the dipolar fields will rotate the magnetization vector $M$ in the
common sense. The opposite is true for edges $\|$~[110]. In the
measurement configurations shown in Fig.~\ref{alignment}(a) and
Fig.~\ref{alignment2}(a) due to small deviations
$\pm\delta\varphi_{H}$ of the field direction from [110] the sense
of the transition can be either CW or CCW. However, independent of
that again the local rotation of $M$ at the [110] edges is opposite
in sense and therefore do not support nucleation in accord with our
experimental results. At higher temperatures $T=24$K and $T=33$K
(Fig.~\ref{domains}(c) and (d)) transitions are CW ($H$-field that
induces the transition is rotated CCW by $15^{\circ}$ from [110]).
Since the easy axis is exactly along [110] it is obvious that
nucleation is again only facilitated at edges $\|$~[1$\bar{1}$0],
where $M$ produces maximum stray fields. However, we believe in this
case the sense of local rotation is not {\it a priori} predictable.
While the model of closure domain formation supports our
experimental observations for fields applied close to [110], the
results in Fig.~\ref{domains2} do not fit into this picture. At low
temperatures for CCW 60$^{\circ}$ DWs we would expect nucleation at
the edges $\|$~[110] and at high temperatures nucleation at those
$\|$~[1$\bar{1}$0]. Instead we observe rather statistical nucleation
within the entire device. We again tend to attribute this difference
in nucleation dynamics to the reduced DW nucleation/propagation
energy $\epsilon_{60^{\circ}}$ with respect to 120$^{\circ}$
DWs~\cite{ourpaper}. Moreover, in the case of Fig.~\ref{domains2} at
low temperatures stray fields proportional to the projection $M\text
{sin}(\theta_{\text{EA}})$ of $M(H=0)$ on [1$\bar{1}$0] are
significantly reduced when compared to those
$M{\text{cos}}(\theta_{\text{EA}})$ in Fig.~\ref{domains}.\newline

It remains to discuss the drastic increase of the number $N(T)$ of
domains involved in the transitions above temperatures $T=23$K. The
problem resembles that of Fatuzzo's domain-nucleation model
developed for ferroelectrics~\cite{Fatuzzo}. As described in the
beginning of this section the increase of $N$ is accompanied by a
change in the average width $w$ of the domains, where $w$ is the
dimension of domains measured along [1$\bar{1}$0] in
Fig.~\ref{domains}. From the time resolved dynamics in our
Kerr-movies above $T=24$K it is evident that after nucleation of a
domain at the Hall-bar edges, DW propagation is mainly happening in
the [110] direction with little change in $w$ of the respective
domain. As proposed by Fatuzzo we therefore attribute the drastic
increase in $N$ to a complex interplay of temperature dependent
nucleation rates $\Gamma(T)$ at the film edges, a reduced DW
mobility $\mu_{[1\bar{1}0]}$ along [1$\bar{1}$0] and effects of
coalescence of domains. If $w_c$ is the average domain width at the
coercive field $H_c$ where $50\%$ of the area of the film has
switched, then the respective number of domains in a given section
of the Hall-bar with a length $l$$\|$~[1$\bar{1}$0] is approximately
$N_c=l/2w_c$. Here $w_c$ will be a function of the mobilities along
$\mu_{[110]}$ and $\mu_{[1\bar{1}0]}$ and the nucleation rates
$\Gamma$. With this we can qualitatively understand the temperature
dependent nucleation dynamics. The Kerr data prove that with
increasing temperature and especially for $T>20$K the ratio between
$\mu_{[110]}$ and $\mu_{[1\bar{1}0]}$ is significantly shifted
towards propagation along [110], which assuming a constant $\Gamma$
would reduce $w_c$ and increase $N_c$. On the other hand we expect
$\Gamma$ to increase with temperature according to a thermally
activated process which supports coalescence of domains at an early
stage after nucleation. Generally, both a decrease in the mobilities
and $\Gamma$ leads to an increase in the coercive field $H_c$ at
constant sweep rates of the magnetic field. Indeed, the temperature
dependence of $H_{c}$ shown in Fig.~\ref{number} (full symbols)
indicates a distinct decrease in slope at $T\approx20$K, which
points towards a change in the mobilities and/or $\Gamma$ ($M$
decreases rather monotonously in this temperature range as shown in
the inset of Fig.~\ref{squid}(a)). Sudden changes in
$\mu_{[1\bar{1}0]}$ or $\Gamma$ would not be unexpected since they
occur in close vicinity to the crossing point between $K_{u}$ and
$K_{c}$ ($\sim$26 Kelvin), where the magnetic transitions change
their character. Above the crossing temperature we interpret the
drastic increase in $N(T)$ to be mainly due to a monotonous
reduction of $\mu_{[1\bar{1}0]}$.

\subsection*{Conclusions}

This work presents an extensive characterization of the temperature
dependent magnetic domain wall dynamics in Hall-bars made from
compressively strained GaMnAs and identifies limits for single
domain wall logic devices in the high temperature regime. Kerr
microscopy allows to locally observe nucleation events of domains as
well as the alignment and propagation behavior of domain walls. A
correlation of the preferential domain wall alignment with respect
to the temperature dependent magnetic easy axis direction is found.
The latter is determined by the temperature dependent in-plane
uniaxial and biaxial anisotropy energy contributions. At low
temperatures magnetically charged domain walls with domain wall
angles considerably smaller than $180^{\circ}$ are observed. Above
the biaxial-to-uniaxial transition temperature this charging effect
is lost and domain walls are oriented along the easy axis. Domain
nucleation is happening almost exclusively at Hall-bar edges aligned
along the [1$\bar{1}$0] uniaxial hard axis direction. This behavior
is attributed to small demagnetizing fields contribution at the
edges of the device, that locally facilitate the magnetic transition
and therefore nucleation of domains. This effect is asymmetric and
favors nucleation at edges $\|$~[1$\bar{1}$0]. This first extensive
study of domain nucleation and propagation dynamics at variable
temperatures in GaMnAs shows that multi-domain states can be avoided
by a suitable device geometry. This together with our finding that
the orientation of domain walls can be tuned by the ratio between
uniaxial and biaxial anisotropy energy has important consequences
for applications in the field of magneto-logics and in particular
for single domain wall devices where domain walls are manipulated
through spin-polarized currents.\newline

\noindent {\bf Acknowledgments}

\noindent We would like to thank Prof. H. Kronm\"uller for valuable
discussions and Ulrike Waizman for conducting the SEM measurements.

\end{document}